\documentclass[iop]{emulateapj}
\usepackage[utf8]{inputenc}

\usepackage{amsmath}
\usepackage{cancel}
\usepackage{bm}
\usepackage{natbib}
\usepackage{graphicx}
\usepackage{footnote}
\usepackage{latexsym}
\usepackage{amssymb}
\usepackage{amsfonts}
\usepackage{hyperref}

\newcommand{\sick}{\texttt{sick}}
\newcommand{\article}{\textit{Article}}
\newcommand{\thetaGrid}{$\bm{\theta_{*}}$}
\def\mean#1{\left< #1 \right>}
\newcommand{\tc}{\textit{The Cannon}}

\begin{document}

\title{\sick, the spectroscopic inference crank}

\author{Andrew~R.~Casey\altaffilmark{1}}

\altaffiltext{1}{Institute of Astronomy, University of Cambridge, Madingley Road,
Cambdridge, CB3 0HA, United Kingdom; \email{arc@ast.cam.ac.uk}}

\begin{abstract}
There exists an inordinate amount of spectral data in both public and private astronomical archives
which remain severely under-utilised. The lack of reliable open-source tools for analysing large volumes of spectra contributes to this situation, which is poised to worsen as
large surveys successively release orders of magnitude more spectra.
In this \article{} I introduce \sick{}, \textit{the spectroscopic inference crank},
a flexible and fast Bayesian tool for inferring astrophysical parameters
from spectra. \sick{} is agnostic to the wavelength coverage, resolving power, or general
data format, allowing any user to easily construct a \textit{generative model}
for their data, regardless of its source. \sick{} can be used to provide a nearest-neighbour
estimate of model parameters, a numerically optimised point estimate, or full Markov Chain Monte Carlo sampling of the posterior probability distributions.
This generality empowers any astronomer to capitalise on the plethora of published synthetic
and observed spectra, and make precise inferences for a host of astrophysical (and nuisance) quantities. 
Model intensities can be reliably approximated from existing grids of synthetic
or observed spectra using linear multi-dimensional interpolation, or a \textit{Cannon}-based model \citep{cannon}. Additional phenomena that transform the data (e.g., redshift,
rotational broadening, continuum, spectral resolution) are incorporated as free parameters
and can be marginalised away.
Outlier pixels (e.g., cosmic rays or poorly modelled regimes) can be treated with 
a Gaussian mixture model, and a noise model is included to account for systematically underestimated variance.
Combining these phenomena into a 
scalar-justified, quantitative model permits precise inferences with credible 
uncertainties on noisy data. 
I describe the common model features, the implementation details, and the default behaviour,
which is balanced to be suitable for most astronomical applications. Using a forward model on low-resolution, high S/N spectra of M67 stars reveals atomic diffusion processes on the order of $0.05$\,dex, previously only measurable with  differential analysis techniques in high-resolution spectra. \sick{} is easy to use, 
well-tested, and freely available online through GitHub under the MIT license. 
\end{abstract}

\section{Introduction}
Most of our understanding of astrophysics has been interpreted from spectra. 
Given how informative spectroscopic data is to our understanding of astrophysics, 
it is not surprising that there has been a substantial increase of 
publicly accessible spectra in the last decade. Large scale surveys have driven this trend, 
each releasing in excess of hundreds of thousands \citep[e.g.,][]{wigglez,boss,
segue,rave,gaia-eso} of spectra. Millions more spectra are expected in the 
coming years \citep[e.g.,][]{lamost, galah}.
 
These spectra are acquired from different astrophysical sources to meet specific 
scientific objectives. They vary in wavelength coverage, resolution, and 
noise distributions. For these reasons many collaborations expend significant 
resources to produce bespoke analysis software for their science program. This usually impedes 
reproducibility, as many codes still remain closed-source nearly 
a decade after the original article was published \citep[e.g.,][]{sspp}, even when the data and results are publicly
accessible. As a consequence any comprehensive literature 
comparison becomes impossible, as systematics are difficult to 
properly characterise without in-depth knowledge of the methods or access to 
the software.

In general there are three types of methods employed for spectral analysis: 
measuring the strengths of spectral features, pure data-generating models or 
template-matching methods. Approaches that measure spectral features \citep[e.g., equivalent widths;][]{daospec} are 
inexpensive, but regularly encounter problems with blended (often hidden) lines 
or continuum placement \citep[e.g., see][for a comparison of techniques]{rodolfo}. In these instances some subjective interaction, tuning, or ad-hoc 
`calibration' is often invoked \citep{kordopatis,sousa}.  Data-generating methods are computationally expensive, repeatedly producing model 
spectra during run-time \citep{sme}. Whilst accurate spectra are produced, the known covariances between stellar parameters are routinely ignored, leading to erroneous results  \citep{torres}. For template-matching methods \citep{matisse,ferre}, synthetic spectra are 
generated once for a subset of 
permutations of astrophysical quantities, usually discretised across a grid.
Although there are differences between 
these methods, the preparatory steps are usually the same: Spectra are placed at 
rest-frame by calculating line-of-sight velocities, typically by 
cross-correlation, before being continuum-normalised to zero and one. 

Credible uncertainties can be difficult to discern from these methods. This is
because the uncertainties in the Doppler-shift, smoothing, sampling, and normalisation steps are 
almost always ignored. These effects result in ill-characterised uncertainties 
in astrophysical parameters. In some cases the uncertainties are simply 
assumed to be approximately the same for all objects.  This is an incorrect 
approach: there are few, if any, examples of homoscedastic datasets in 
astrophysics. The noise properties of each spectrum \textit{are} different, and 
the parameter uncertainties (random and systematic) will differ for every object. 
Consequently the uncertainties in astrophysical parameters by template-matching 
methods are generally found to either be incorrectly assumed, under-estimated, 
or at least ill-characterised.

In addition to affecting the uncertainties, the effects of redshift, continuum 
normalisation and smoothing \textit{will} further bias the reported 
maximum-likelihood parameters. For example, experienced stellar spectroscopists will frequently 
differ in their decision of continuum placement even in the most 
straightforward cases (e.g., metal-poor stars). There are a number 
of controversial examples within the literature where the subjective (human) 
decision of continuum placement have significantly altered the scientific 
conclusions \citep[e.g., see][where this issue is discussed in great 
detail]{kerzendorf}. The implications of these phenomena \textit{must} be 
considered if we are to understand subtle astrophysical processes. Spectroscopy 
requires open(-sourced) objectivity. One should endeavour to incorporate these phenomena as 
free parameters into a generative model and infer them simultaneously with the 
astrophysical parameters.

In this \article{} I present \sick{}: a flexible, well-tested, MIT-licensed probabilistic 
software package for inferring astrophysical (and nuisance) quantities from spectroscopic data. \sick{} employs  
approximations to data-generating models. Instead of modelling expensive astrophysical processes (e.g., stars, supernova, 
and any other interesting astrophysical processes) at run-time, \sick{}
approximates the model intensities from pre-computed grids/sets of spectra. Contributory phenomena (e.g., continuum, redshift, rotational broadening, spectral resolution) are included as free
parameters within an objective scalar-justified model. This approach is suitable 
for a plethora of different astrophysical 
processes, allowing any user to easily construct a generative model using existing grids of
published spectra, and infer astrophysical properties from their data. Aspects of 
the probabilistic model are described in Section \ref{sec:model}. The analysis methodology
is discussed in Section \ref{sec:method}, and a toy model is presented in Section \ref{sec:toy-model}. In Section \ref{sec:utility} I present a suitable scientific application that demonstrates the power of forward models with existing data, instead of existing subjective approaches. I conclude in Section \ref{sec:conclusion} with references to 
the online documentation and applicability of the software.

\section{The Generative Model}
\label{sec:model}

The generative model described below is agnostic as to \textit{what} the 
astrophysical parameters actually describe\footnote{However given the research background of the author, these examples will focus on stellar applications.}. Typical examples might be properties of supernova (e.g., explosion energies and luminosities), galaxy characteristics from integrated light, or mean plasma properties of a stellar photosphere. There 
are a plethora of spectral libraries (observed and synthetic) published for these 
types of applications \citep[e.g.,][]{snid,pegase,phoenix,pollux}. All of these 
models are fully-\sick{} compatible\footnote{Libraries of published spectra that are ready-to-use are available for download through the \texttt{sick download} command line tool.}. 

In an ideal world one would avoid spectrum approximations entirely and aim to solve the hydrodynamic and radiative transfer equations to produce accurate model spectra in real-time. This would certainly be a computationally expensive endeavour. It may also be unnecessary. In practice \sick{} can be sub-classed (by inheriting from the \texttt{BaseModel} class) to produce more realistic spectra at run-time or to allow approximations within existing grids with different approaches. The code is designed to be flexible to suit a range of scientific objectives.


Let us assume that there exists a set of astrophysical parameters \thetaGrid{}
that I wish to infer from some data. I first must produce a spectrum of normalised model intensities (e.g., between 0 and 1) from an
existing set (or grid) of spectra. The grid of spectra are expected to span a suitable range of
\thetaGrid{} values, but are not required to be regularly spaced in \thetaGrid{}. Indeed, high-quality observed spectra with irregularly-spaced
yet precisely-measured values of \thetaGrid{} are perfectly acceptable. There are currently two techniques for producing intensities at any \thetaGrid{} value in \sick{}, which are described in the following sections.

\subsection{$N$-Dimensional Linear Interpolation}
\label{sec:model-interpolation}
The simplest available method is to linearly interpolate between the grid of model
spectra. The detailed features of most astrophysical model spectra are unlikely to be well-captured by crude linear interpolation in high dimensions. However, the $N$-dimensional linear interpolation scheme is efficient and may be suitable for many astrophysical instances where only a point-estimate of \thetaGrid{} is required. 

This approach uses the Quickhull \citep{quickhull} algorithm in SciPy \citep[\texttt{scipy.interpolate.griddata} and \texttt{scipy.interpolate.LinearNDInterpolator};][]{scipy} to interpolate in $N$-dimensions. The convex hulls required for Quickhull can be globally produced for the entire grid, or for \texttt{N}$_{\rm grid}$ local points surrounding the initial estimate $\bm{\theta_{*,{\rm estimate}}}$ (see Section \ref{sec:method-initial-estimate}), where \texttt{N}$_{\rm grid}$ can be specified in the model configuration file. Quickhull relies  on Voronoi tessellation, which produces extremely skewed cells when the grid points $\{\bm{\theta_*}\}^{\rm grid}$ vary significantly in magnitude (e.g., as $T_{\rm eff}$ does with respect to $\log{g}$ in the examples presented here) and will introduce substantial errors in the interpolation. To minimise these effects, \sick{} automatically scales the grid values $\{\bm{\theta_*}\}^{\rm grid}$ (in both global and local scenarios) to make $\delta\theta_{*,dim}$ approximately equal (e.g., $\delta{}T_{\rm eff,scaled} \approx \delta{}\log{g}_{scaled}$).

The model intensities $I_{\lambda,m}\left(\bm{\theta_*}\right)$ at wavelengths $\bm{\lambda_m}$ for some arbitrary \thetaGrid{} can be approximated by interpolating from a `nearby' (local or global) grid $\{\bm{\theta_*}\}^{\rm grid}$,

\begin{equation}
I_{\lambda,m}\left(\{\bm{\theta_*}\}^{\rm grid}\right) \leadsto I_{\lambda,m}\left(\bm{\theta_*}\right),
\end{equation}

\noindent{}where, following the nomenclature in parallel work by \citet{czekala}, I denote the symbol $\leadsto$ as an interpolation operator.

In practice large grids of model spectra are automatically cached using efficient memory-mapping, allowing for the total size of the model grid to far exceed the available random access memory. In other words, grids of model spectra that are hundreds of Gb in size can be efficiently accessed from an external hard disk on any reasonably modern CPU.

\subsection{The Cannon}
\label{sec:model-cannon}

\tc{} \citep{cannon} is a data-driven approach for stellar label determination. The approach makes use of a training set of stars where labels (i.e., $\bm{\theta_*}$) have been determined with high fidelity. The normalised, rest-frame pixel information (on a common binning scale) is then used to build a spectral model for a larger sample of stars. \citet{cannon} use observed spectra from APOGEE \citep{apogee} as their training set, and project spectra from the entire survey to efficiently determine stellar parameters. Here I employ \tc{} as a method for producing model intensities within a grid, using synthetic labels.

For the full description of \tc{}, the reader is directed to \citet{cannon}. Here I describe the algorithm in brief, with a focus on the implementation details in \sick{}. A \textit{Cannon} model is characterised by a coefficient vector $\bm{\theta}_{\lambda,m}$ that allows for the prediction of the intensitiy $I_{\lambda,m}$ for a given label vector $\bm{l}$, such that at the $i$-th model pixel with wavelength $\lambda_{m,i}$:

\begin{equation}
I_{\lambda,m,i} = g(\bm{\lambda}_{m,i}|\bm{\theta}_{\lambda,m,i}) + {\rm noise}
\end{equation}

The vector $\bm{\theta}_{\lambda,m,i}$ is a set of spectral model coefficients at each $\lambda_{m,i}$. Thus for the $i$-th pixel there are a number of coefficients $\bm{\theta_{\lambda,m,i}}$ and a scatter term $s_{\lambda,m,i}$ which we must solve (train). The complexity in $\bm{\theta_{\lambda,m}}$ can be adjusted: a linear model is the simplest option, whereas \citep{cannon} employ a quadratic-in-labels model with linear cross-terms. If none is specified, \sick{} will default to a quadratic-in-labels model with linear cross-terms. The label vector $\bm{l}_m$ (linear vector shown),

\begin{equation}
\bm{l}_m \equiv [1, \theta_{*,i} - \overline{\theta_{*,i}},\theta_{*,j} - \overline{\theta_{*,j}}, \cdots{}]
\end{equation}

\noindent{}is generated from a human-readable description of the label vector. The $\overline{\theta_{*}}$ offsets are taken as the means of the training set. Before using \tc{}, we need to solve for the coefficients $\{\bm{\theta}_{\lambda,m},s_{\lambda,m}\}$ for the label vector $\bm{l}_{\lambda,m}$ by optimising the log-likelihood function for each $\lambda_{m,i}$ pixel,

\begin{multline}
\label{eq:cannon-likelihood}
 \ln{}p(I_{\lambda,m,i}|\bm{\theta}_{\lambda,m}^\intercal{},\bm{l}_{\lambda,m,i},s_{\lambda,m,i}^2) = \cdots{} \\
 \left. -\frac{1}{2}\frac{[I_{\lambda,m,i} - \bm{\theta}_{\lambda,m}^{\intercal}\cdot{}\bm{l_{\lambda,m,i}}]^2}{s_{\lambda,m,i}^2 + \sigma_{\theta,*,i}^2} -\frac{1}{2}\ln(s_{\lambda,m,i}^2 + \sigma_{\theta,*,i}^2)\right.
\end{multline}

\noindent{}where $\sigma_{\theta_{*,i}}$ is the intensity uncertainty in the $i$-th pixel of the model grid point with parameters $\bm{\theta_{*}}$. For the remainder of this \article{} I will only use synthesised spectra to produce $I_{\lambda,m}(\bm{\theta_*})$, and thus $\sigma_{\theta_{*,i}}$ is zero. However, this term is included in Equation \ref{eq:cannon-likelihood} to demonstrate that \tc{} implementation in \sick{} can be trained from existing observed spectra, as originally utilised by \citet{cannon}. \\

\subsection{Ancillary Effects}
\label{sec:ancillary-effects}

With the model intensities $I_{\lambda,m}(\bm{\theta_*})$ produced, there are still a number of additional phenomena that must be considered before the model can be reliably compared to the data. I will describe the dominant effects in the context of a single observed channel (order/beam/aperture). However, \sick{} allows the user to model any combination of the effects described below, with flexible options for handling multiple channels. For example, although redshift $z$ is an astrophysical phenomena, if some channels do not benefit from telluric absorption or imprints of Earth-bound rest-frame spectra in their observations, then the wavelength calibration will rely on different arc lines, and consequently be slightly varied in each channel. The user can opt for a single parameter $z$, or implement separate redshift parameters per channel, with an optional strong joint prior on those parameters.

The model intensities $I_{\lambda,m}(\bm{\theta_*})$ that have been generated should always be of higher spectral resolution than the data. 
Therefore it is necessary to convolve the model spectra $\{\bm{\lambda_m},\bm{I_{\lambda,m}, \bm{s_{\lambda,m}}}\}$. For these reasons I introduce the spectral resolution $\mathcal{R} = \frac{\lambda}{\Delta\lambda}$ as an additional nuisance parameter. The resolving power can be modeled as a single resolution for all orders, or through additional parameters for each order. For stellar applications, the observed rotational velocity $v\sin{i}$ \citep{gray} can be modelled by an additional kernel, convolved with the $\mathcal{R}$ kernel.

The convolution, resampling, and redshift steps are performed simultaneously in
\sick{}. Given the model wavelengths $\bm{\lambda_m}$ and observed  wavelengths $\bm{\lambda_{o}}$, \sick{} efficiently constructs a sparse matrix $\bm{S}$ of size $N_{\lambda,m} \times N_{\lambda,o}$ such that the expected intensity at an observed pixel $E_{\lambda_{o}}$ is given by the dot product of $I_{\lambda,m}(\bm{\theta_*})$ and $\bm{S}(\bm{\theta_s})$,:

\begin{equation}
E_{\lambda,o}(\bm{\theta_*},\bm{\theta_s}) = I_{\lambda,m}(\bm{\theta_*}) \cdot \bm{S}(v\sin{i},\bm{z},\bm{\mathcal{R}},\bm{\lambda_m},\bm{\lambda_o}).
\end{equation}

\noindent{}where $\bm{\theta_{s}} = [v\sin{i},\bm{z},\bm{\mathcal{R}},\bm{\lambda_m},\bm{\lambda_o}]$. The vectors $\bm{z}$ and $\bm{\mathcal{R}}$ represent different redshifts and resolving powers in multiple channels.

\begin{figure*}[t!]
\includegraphics[width=\textwidth]{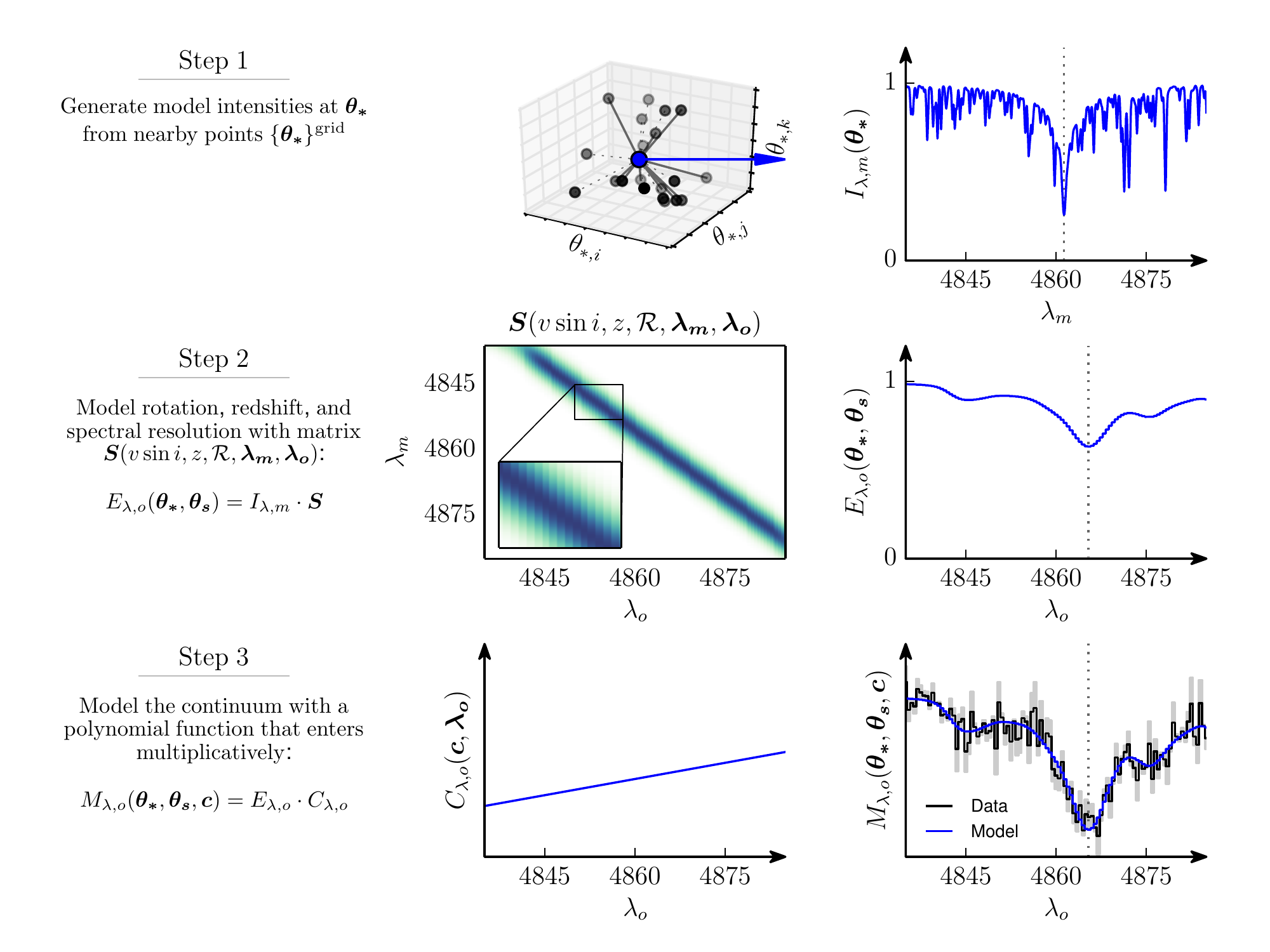}
\caption{The steps required to produce the model spectrum $M_{\lambda,o}(\bm{\theta_*},\bm{\theta_s},\bm{c})$. This figure is intended to clarify the mathematical nomenclature and visualise the data-generating procedure. Transformations occur from left-to-right, starting from the top row. The spectrum in the top right is a high-resolution ($\mathcal{R} \sim{} 20,000$) portion surrounding H$\alpha$, which is marked in all panels in the rightmost column. The inset axes in the center panel shows a $5\times5$\,{\AA} zoom-in of the convolution kernel. Redder wavelengths have a larger kernel width due to $\mathcal{R}$. Positive redshifts move the position of the diagonal convolution to the right. The final model spectrum is shown compared to a low-resolution ($\mathcal{R} \sim 2,000$) spectrum of a solar-like star with S/N $\sim{}30$ per pixel. The additional nuisance parameters $f$, $p_b$, $v_b$ only contribute during the calculation of the likelihood $\mathcal{L}$, and do not affect the data generated by the model.}
\label{fig:data-generation-illustration}
\end{figure*}

This data-generation procedure (as illustrated in Figure \ref{fig:data-generation-illustration}) accounts for all convolutions ascribed above and ensures the total flux is preserved: in notation $\bm{S}_{i,j}$, $\sum_{j=0}^{N_{\lambda,m}} S_{i,j} = 1$ for all $i$. The total width of the convolution kernel (the pixel-convolved line spread function) across the matrix $\bm{S}$ for a single channel is dependent on $v\sin{i}$ and $\mathcal{R}$ (and subsequently $\bm{\lambda_o}$). The $(x,y)$ pixel centroid of the convolution kernel along any row or column depends on the model wavelengths $\bm{\lambda_m}$ and the redshift $z$ (through $\lambda_{shifted} = \lambda_{m}(1 + z)$), as well as the wavelengths of the observed pixel $\bm{\lambda_o}$. For situations where no kernel convolution is required, a comparable $\bm{S}$ matrix is produced to rebin the model to the data, given a redshift $z$.

The construction of $\bm{S}$ constitutes a non-negligible component to the total computational budget for each probability evaluation. For this reason a few (optional) approximations have been implemented to minimise this cost. A least-recently used cacher is employed by default to minimise the number of matrix constructions. This uses a small portion of random access memory to retain the \texttt{N}$_{\rm \texttt{LRU}}$ most common sets of $\{\bm{\theta_s},\bm{S}\}$.
If $\bm{S}(\bm{\theta_s} + \delta\bm{\theta_s})$ is required and $\delta\bm{\theta_s}$ is sufficiently small (i.e., below some prescribed tolerance) such that 
$\bm{S}(\bm{\theta_s}) \approx \bm{S}(\bm{\theta_s} + \delta\bm{\theta_s})$, then future calls of 
$\bm{S}$ within the range $\bm{\theta_{s}} \pm \delta\bm{\theta_s}$ will return the previously 
calculated matrix $\bm{S}$ instead of reconstructing it. The tolerances are configurable, and 
default to the sub-km\,s$^{-1}$ level for $v\sin{i}$ and $z$, and $\delta\mathcal{R} < 1$. 
Alternatively, the construction of $\bm{S}$ can be completely avoided by approximating the 
convolution with a single kernel width $\sigma$ at all $\bm{\lambda_o}$, where an interpolation routine is used to calculate the expected intensity 
$E_{\lambda,o}(\bm{\theta_*},\bm{\theta_s})$ at $\bm{\lambda_o}$. The extent of 
approximation that can be afforded will vary on the scientific objectives, but the default 
behaviour in \sick{} is balanced to be computationally efficient, and suitable for most scientific applications.

I have produced the expected intensities $E_{\lambda,o}(\bm{\theta_*,\theta_s})$ for some arbitrary \thetaGrid{} that can fairly represent any rotational broadening of the source, the object's redshift, as well as the resolving power of the instrument and the location of the CCD pixels. However, these intensities are not representative of real world data. 
The height (i.e., photon counts) and shape of an observed
spectrum is a function of the source magnitude, exposure time, instrument 
sensitivities, atmospheric conditions, interstellar extinction, 
and a host of unaddressed effects. 

For these reasons a function $C_{\lambda,o}(\bm{c})$ is required to normalise\footnote{The normalisation 
process is frequently abused by stellar spectroscopists in the literature. 
Wherever possible, data should not be transformed. One should seek to fit a 
model to the data, not the other way around.} the model to the data.
Although the function $C_{\lambda,o}$ incorporates a number of effects (e.g., source 
blackbody temperature, dust, instrument sensitivities), they are phenomena that 
usually cannot be separated without additional information, and here I only 
care about their combined effect. The continuum is modeled as a polynomial
that enters multiplicatively,

\begin{equation}
\label{eq:continuum}
    C_{\lambda,o}(\bm{c}) = \sum_{i=0}^{j}c_{i}\bm{\lambda_o}^{i}
\end{equation}

\noindent{}where the maximum polynomial degree is specified by the user. For data spanning
multiple channels, I denote $\{\bm{c}\}$ to represent the continuum coefficients $\bm{c}$ in each observed channel. For brevity I define the expected model fluxes at the observed pixels with wavelengths $\bm{\lambda_{o}}$ as:

\begin{equation}
    M_{\lambda,o}(\bm{\theta_*},\bm{\theta_s},\bm{c}) = E_{\lambda,o}(\bm{\theta_*},\bm{\theta_s})\cdot{}C_{\lambda,o}(\bm{c},\bm{\lambda_o})
\end{equation}

Thus, the model intensities $I_{\lambda,m}(\bm{\theta_*})$ at model wavelengths $\bm{\lambda_m}$ are normalised between 0 and 1 and have no units, as are the expected intensities $E_{\lambda_o}(\bm{\theta_*},\bm{\theta_s})$ at observed wavelengths $\bm{\lambda_{o}}$. However, the model fluxes $M_{\lambda_o}(\bm{\theta_*},\bm{\theta_s},\bm{c})$ at observed wavelengths $\bm{\lambda_o}$ will have the same `units' as the observations, be it in photon counts (i.e., arbitrary units) or energy flux density (erg\,s$^{-1}$\,cm$^{-2}$\,\AA{}$^{-1}$). That is to say, \sick{} does not require the data to be flux-calibrated.

\subsection{Underestimated Variance}
\label{sec:model-underestimated-variance}

For data generating models that incorporate an uncertainty in producing each pixel value $\bm{s_{\lambda,m}}$ (e.g., Section \ref{sec:model-cannon}), the total variance $\bm{s_{\lambda,o}}^2$ in each pixel is given by:

\begin{equation}
\bm{s_{\lambda,o}}^2 = \bm{\sigma_{\lambda,o}}^2 + \bm{C_{\lambda,o}}^2(\bm{s_{\lambda,m}}\cdot{}\bm{S})^2.
\end{equation}

The $N$-dimensional linear interpolation method in Section \ref{sec:model-interpolation} does not allow for a $\bm{s_{\lambda,m}}$ term and thus $C_{\lambda,o}^{2}(\bm{s_{\lambda,m}}\cdot{}\bm{S})^2$ cancels to zero for the linear interpolation method. However, irrespective of the data-generating method employed, the observed pixel uncertainties $\bm{\sigma_{\lambda,o}}$ are usually under-estimated. This may be due to unpropagated uncertainties during data reduction, and/or more commonly a result of untreated covariance between neighbouring spectrograph pixels. Ideally the full covariance matrix $\bm{\Sigma}$ should be used to fit the data, however having access to the proper covariance matrix from a data reduction pipeline is not a common scenario.

A robust approach would be to model the neighbouring covariance along the diagonal of $\bm{\Sigma}$ with a Gaussian process \citep[e.g., see excellent work by][]{czekala}. The downside to employing a Gaussian process to model the global covariance is that matrices of considerable size require regular inversions, potentially adding considerable cost to the computational budget. A simpler approach is to assume the pixel uncertainties $\bm{\sigma_{\lambda,o}}$ are systematically underestimated by some fractional amount $f$. In this scenario the observed variance $\bm{\sigma_{\lambda,o}^2}$ for a given pixel is given by:

\begin{equation}
\bm{s_{\lambda,o}}^2 = \bm{\sigma_{\lambda,o}}^2 + \bm{C_{\lambda,o}}^{2}(\bm{s_{\lambda,m}}\cdot{}\bm{S})^2 + f^{2}\bm{M_{\lambda,o}}^{2}
\label{eq:total-variance}
\end{equation}

\subsection{Outliers}
\label{sec:model-outliers}

I now consider the handling of outliers in the data. These may be in the form 
of cosmic ray spikes, improper calibration of the data, telluric features, or simply 
poorly modelled spectral regimes. These pixels can be treated in two ways within
\sick{}: a Gaussian mixture of the spectrum model and an outlier model, or with
semi-constrained $\sigma$-clipping at run-time. When a mixture model is employed, the data 
are fit by the sum of amplitudes ($1 - p_b$ and $p_b$, respectively, where $b$ represents the `background' outlier model) of two 
distributions: the model fluxes $\bm{M_{\lambda,o}}$, and a normal distribution with mean  $\bm{C_{\lambda,o}}$ and additional variance $v_b$ such that the total variance in a given pixel for the outlier model is $\bm{s_{\lambda,o}}^2 + v_b$.

This requires the 
inclusion of two additional nuisance parameters: $p_b$ and $v_b$. 
The prior $p(p_b) = \mathcal{U}(0, 1)$ is hard-coded in \sick{}.
The prior 
distribution function $p(v_b)$ is similarly fixed, requiring $v_b$ to always be positive 
(Equation \ref{eq:default_priors}), and as such the outlier distribution will 
\textit{always} have a larger variance.  Distributions of smaller variance are more 
informative, so conceptually a fit to the model fluxes $\bm{M_{\lambda,o}}$ is generally 
preferred wherever possible.

\subsection{Priors}
\label{sec:model-priors}

Priors represent our initial knowledge about a particular parameter before 
looking at the data, and are necessary for any Bayesian analysis.
A number of 
different prior distributions can be specified by the user in the \sick{} model 
configuration file. The following uninformative prior distributions are assumed (for all channels,
where appropriate) unless otherwise specified:

\begin{eqnarray}
p\left(\theta_{*,dim}\right) &\,=\,& \mathcal{U}\left(\min\left[\{\bm{\theta_*}\}^{\rm grid}_{dim}\right], \max\left[\{\bm{\theta_*}\}^{\rm grid}_{dim}\right]\right) \\
p\left(z,\{\bm{c}\}\right) &\,=\,& 1 \\
p\left(\ln{f}\right) &\,=\,& \mathcal{U}\left(-10, 1\right) \\
p\left(p_b\right) &\,=\,& \mathcal{U}\left(0, 1\right) \\
p\left(\mathcal{R},v_b\right) &\,=\,& \left\{
\begin{array}{c l}      
    1\,, &\mbox{for values greater than zero}\\
    0\,, &\mbox{otherwise}
\end{array}\right.
\label{eq:default_priors}
\end{eqnarray}

There are some subtleties to enforcing $p(\bm{\theta_*})$. A consequence of allowing irregular model grids in the $N$-dimensional linear interpolation model is that occasionally a spectrum cannot be interpolated for some $\bm{\theta_*}$, even if it is bound within $\left(\min\left[\{\bm{\theta_*}\}^{\rm grid}_{dim}\right], \max\left[\{\bm{\theta_*}\}^{\rm grid}_{dim}\right]\right)$ because it will be outside the convex hull. In these cases $p(\bm{\theta_*}) = 0$ and thus $\ln{\mathcal{P}} = -\infty$. On the other hand if a \textit{Cannon} spectral model is used, data can be produced beyond the strict parameter limits that make up the training set. Although this is not a severe restriction on \textit{The Cannon}, by default \sick{} is cautious and enforces  $p(\bm{\theta_*}) = 0$ for points outside the limits of the training set. This option can be disabled by the user.

It is clear that the priors on $\{\bm{c}\}$ intuitively should not be uniform. Higher order terms of a polynomial sequence should have much smaller priors, as their absolute magnitudes are expected to be much smaller than lower order terms. In practice the initialization of continuum parameters is normally sufficient such that the default uniform priors on $\{\bm{c}\}$ pose no problem, but for the expert user there is clearly room for a formal, well-founded prior to be enforced on $\{\bm{c}\}$, which can be enabled in the \sick{} model description. Finally, negative generated model fluxes $M_{\lambda,o}$ are considered unphysical, and set to non-finite values.

\subsection{The Likelihood Function}
\label{sec:model-likelihood}

The likelihood function has an additional term if outlier pixels are treated with a mixture model. The parameter $\bm{\Theta} \equiv [\bm{\theta_*},v\sin{i},\bm{R},\bm{z},\{\bm{c}\},\bm{f}]$ describes all parameters in the model. If a \textit{Cannon} model is used, the implication is that the parameters $\{\bm{\theta_{\lambda,m}},\bm{s_{\lambda,m}}\}$ have been solved for at each pixel, and are thus already folded in to $I_{\lambda,m}(\bm{\theta_*})$ and $\bm{s_{\lambda,o}}^2$. Similarly, due to the convolution and binning matrix $\bm{S}$, the model flux at a given pixel $\lambda_{o,i}$ is reliant on knowing the neighbouring model and observed wavelengths $\bm{\lambda_m}$ and $\bm{\lambda_o}$, but for brevity I do not explicitly specify these terms in Equation \ref{eq:p_model} below.

I can now describe the frequency (or probability distribution) for the flux at each pixel (of $\{\lambda_{o},F_{\lambda,o},\sigma_{\lambda,o}\}$) $p(\bm{F_{\lambda,o}}|\bm{\lambda_o},\bm{\sigma_{\lambda,o}},\bm{\Theta})$ for the observed data $\bm{F_{\lambda,o}}$:

\begin{equation}
p\left(\bm{F_{\lambda,o}}|\bm{\lambda_{o}},\bm{\sigma_{\lambda,o}},\bm{\Theta}\right) = \frac{1}{\sqrt{2\pi\bm{s_{\lambda,o}}^2}}\exp{\left(-\frac{\left[\bm{F_{\lambda,o}} - \bm{M_{\lambda,o}}\right]^2}{2\bm{s_{\lambda,o}}^2}\right)}
 \label{eq:p_model}
\end{equation}

\noindent{}where $\bm{s_{\lambda,o}}^2$ was defined in Equation \ref{eq:total-variance}. If the outlier pixels are also being modelled then the probability distribution for $\bm{F_{\lambda,o}}$ becomes a mixture of two models:

 \begin{multline}
p\left(\bm{F_{\lambda,o}}|\bm{\lambda_{o}},\bm{\sigma_{\lambda,o}},\bm{\Theta},p_b,v_b\right) = \cdots{} \\
(1 - p_b) \times p\left(\bm{F_{\lambda,o}}|\bm{\lambda_{o}},\bm{\sigma_{\lambda,o}},\bm{\Theta}\right) + \cdots{} \\
\left. p_{b} \times p_{background}\left(\bm{F_{\lambda,o}}|\bm{\lambda_{o}},\bm{\sigma_{\lambda,o}},\bm{\Theta},v_b\right) \right.\\
\end{multline}

\noindent{}Where $p_{background}$ is defined as:

\begin{multline}
p_{background}\left(\bm{F_{\lambda,o}}|\bm{\lambda_{o}},\bm{\sigma_{\lambda,o}},\bm{\Theta},v_b\right) = \cdots{} \\
\left. \frac{1}{\sqrt{2\pi{}(\bm{s_{\lambda,o}}^2 + v_b)}}\exp{\left(-\frac{[\bm{F_{\lambda,o}} - \bm{C_{\lambda,o}}]^2}{2[\bm{s_{\lambda,o}}^2 + v_b]}\right)}\right.
\end{multline}

\section{Methodology}
\label{sec:method}

\sick{} aims to be flexible to achieve different scientific objectives, depending on the data volume and the computing resources available. Generally, when I have acquired some data, I seek either a coarse guess of the model parameters $\bm{\Theta}$, a numerically-optimised point estimate of $\bm{\Theta}$ (no uncertainties), or full sampling of the posterior probability distribution. 
\sick{} has three primary analysis functions to suit these scenarios. Below I list the abridged \sick{} command line usage for each situation, as well as a brief description:

\begin{enumerate}
\item \texttt{sick estimate <model> <data>}\\An initial estimate of the model parameters $\bm{\Theta}_{estimate}$ is obtained by cross-correlating a pseudo-normalised copy of the data against the entire model grid (or some subset thereof; Section \ref{sec:method-initial-estimate}). The nearest neighbour $\bm{\theta_*}^{\rm nearest}$ is returned, and $\{\bm{c}\}$ and $\bm{z}$ are estimated from $\bm{\theta_*}^{\rm nearest}$. 

\item \texttt{sick optimise <model> <data>}\\Numerical optimisation of $-\ln{\mathcal{P}}$ begins from $\bm{\Theta}_{estimate}$, unless an initial guess is provided. If no minimisation algorithm is selected, a scaled and bounded version of the Broyden, Fletcher, Goldfarb, and Shanno (BFGS; Section \ref{sec:method-optimise}) algorithm is used.

\item \texttt{sick infer <model> <data>}\\Markov Chain Monte Carlo (MCMC) sampling begins from the numerically optimised point $\bm{\Theta}_{estimate}$. Sampling occurs for at least $2,000$ steps with 200 walkers ($4\times10^5$ probability evaluations), with convergence automatically determined from the autocorrelation functions, unless conflicting sampling requirements have been provided by the (presumably expert) user.
\end{enumerate}

\subsection{Initial Estimate}
\label{sec:method-initial-estimate}

I require a good initial estimate of the model parameters $\bm{\Theta}_{estimate}$. Initially \sick{} fits each observed channel with a polynomial (with degree set by the user in the model configuration file), discards pixels that deviate by more than 4$\sigma$, and repeats the fit. A copy of the data is divided by the fitted continuum, yielding a `pseudo-normalised' spectrum (e.g., the spectrum is `normalised' without any consideration of strong molecular bands or continuous opacities depressing the entire spectrum)\footnote{This procedure constitutes the antithesis of an earlier footnote, but is only being performed to facilitate a cheap comparison between the data and all possible models.}.
This `pseudo-normalised' spectrum is then cross-correlated against the entire  model grid, or \texttt{N}$_{\rm{\texttt{grid\_estimate}}}$ equispaced points across $\{\bm{\theta_*}\}^{\rm grid}$.

The relative peak of the cross-correlation function (CCF) $\mathcal{F}_{mo}$ provides a reliable metric of similarity between two spectra, thereby providing a cheap estimate of the model parameters $\bm{\theta_*}$. Given the nearest-neighbour guess of $\bm{\theta_*}$ and the redshift $z$, the optimal continuum coefficients $\bm{c}$ can then be calculated algebraically after pseudo-sampling (i.e., interpolating) $\bm{I_{\lambda,m(1+z)}}$ on to $\bm{\lambda_{o}}$:

\begin{equation}
\bm{F_{\lambda,o}} \approx I_{\lambda,m(1 + z)}(\bm{\theta_*}) \cdot{} C_{\lambda_{o}}(\bm{c}, \bm{\lambda_{o}})
\label{eq:estimate-continuum}
\end{equation}

When multiple channels are present (and the redshift is being modelled
separately in each channel), an estimate of the closest model parameters $\bm{\theta_*}$ is provided by each channel.
In this scenario \sick{} calculates the optimal continuum coefficients $\bm{c}$ in
each channel, for each unique value of the set $\{\bm{\theta_{*}}\}$ returned from the peaks of the CCFs.

The $\chi^2$ difference (calculated using all channels)
between the approximate model (Equation \ref{eq:estimate-continuum}) and the data is calculated for each entry in $\bm{\theta_*}$. The point with the lowest total $\chi^2$ value is taken as the initial estimate of $\bm{\theta_*}$ (and its corresponding redshift(s) $\bm{z}$ and continuum coefficients $\{\bm{c}\}$). The spectral resolution(s) $\bm{\mathcal{R}}$ are estimated from the wavelength spacing $\frac{\lambda}{\Delta\lambda}$ in each channel. The fraction of underestimated variance is assumed to be high ($\ln{\bm{f}} = 0.5$), and the outlier fraction is initially estimated to be small (1\%), with the outlier variance assumed to be comparable to the observed pixel variance $v_{b} = \mean{\bm{\sigma_{\lambda,o}}}^2$.

\subsection{Optimisation}
\label{sec:method-optimise}

The model parameters $\bm{\Theta}$ are then numerically optimised, using $\bm{\Theta}_{estimate}$ as the starting point. I numerically optimise the 
parameters $\bm{\Theta}$ by minimising the negative log-probability 
$-\ln{\left(\mathcal{P}\right)}$. A number of suitable minimisation 
algorithms are available in \sick{} through the SciPy \citep{scipy} optimization 
module:

\begin{itemize}
\item BFGS \citep{byrd_1995,zhu_1997,morales_2011,nocedal_wright} [default]
\item Modified Powell's method \citep{powell_1964,press_2002}
\item Non-linear conjugate gradient method \citep{nocedal_wright}
\item Truncated Newton conjugate-gradient method \citep{nash_1984, nocedal_wright} 
\item Nelder-Mead \citep{nelder-mead}
\end{itemize}

 If bounded information is available (e.g., boundaries of $\{\bm{\theta_*}^{\rm grid}\}$ or limits from uniform priors), \sick{} will use constrained implementations of the algorithms above, where they exist.
For algorithms that utilise a single parameter scaling factor for gauging convergence (e.g., \texttt{factr} in BFGS), \sick{} automatically scales $\bm{\Theta}$ to place the parameters in the same order of magnitude. Tunable convergence parameters for each optimisation algorithm are also configurable through the model configuration file. However, the default options in \sick{} should be suitable for most purposes.

Model parameters can also be optionally fixed during the optimisation process. For example, if only a point estimate of $\bm{\Theta}$ is required (e.g., no MCMC sampling) and the procedure in Section \ref{sec:method-initial-estimate} provides a reliable measure of $\bm{z}$, one might choose to keep $\bm{z}$ fixed for the optimisation process and solve for the remaining $\bm{\Theta}$.

\subsection{Monte-Carlo Markov Chain Sampling}
\label{sec:method-mcmc}

The aforementioned steps efficiently provide an 
accurate point estimate of the optimal parameters $\bm{\Theta}_{optimised}$. 
However they do not provide a measure of uncertainty on $\bm{\Theta}$, which
is usually more important than a single value\footnote{A useful analogy to emphasise the importance of uncertainties is a hypothetical scenario where you are told a measure of some unfamiliar physical object, without being told the \textit{unit} of measure. Without knowing anything about the object (e.g., rough size, its purpose, etc), the measure could be miles, volume in mm$^{3}$, temperature -- you don't know! Similarly if I measured some astrophysical quantity (one which could be expressed in a variety of units) to be $X$ -- it's equally uninformative to omit the uncertainties as it is to omit the units! There is no information about the \textit{scale} or \textit{variance} of $X$. Uncertainties are important.}. \sick{} employs the affine-invariant ensemble Metropolis-Hastings sampler proposed by \citet{goodman;weare} and implemented by \citet{emcee}. The initial distribution of $\bm{\Theta}$ (hereafter called the initial state $\alpha$) are drawn from a small multi-dimensional ball around the optimised parameters $\bm{\Theta}_{optimised}$.

Before the posterior distributions of $\bm{\Theta}$ can be properly sampled, the MCMC chains must be thermalised (i.e., `burnt in') from the initial proposal $\alpha$ to the equilibrium distribution $\pi$. It can be shown \citep[e.g.,][]{chung_2006} that a Markov chain will converge to $\pi$ as ${t \rightarrow\infty}$. However the initial transit period from $\alpha$ to $\pi$ must be discarded; our posterior $p(\bm{\Theta})$ should not depend on the initial distribution $\alpha$. Ensuring a Markov chain converges to equilibrium in finite time is a fundamental topic in statistics. 

In an attempt to make \sick{} easy to use, some default behaviour has been introduced to routinely evaluate whether convergence has \textit{probably} been achieved. Consider the transit from the $\alpha$ to $\pi$ distributions. Each successive state ($\bm{\Theta}_o$, $\bm{\Theta}_1$, $\ldots{}$) of the Markov chains are correlated. In other words, each state depends slightly on the previous state. This is the exact opposite of what I actually want to achieve, as I seek $\pi$ to be effectively independent of the initial distribution $\alpha$. Thus the auto-correlation between successive states is informative of -- amongst other things -- whether the Markov chains are near $\alpha$ or $\pi$. If one considers an observable $f$ (e.g., any parameter in $\bm{\Theta}$), I can estimate its mean,

\begin{equation}
\mean{f} = \int{}f(\phi)p(\phi)d\phi
\end{equation}

\noindent{}where $p(\phi)$ is the probability density function. The unnormalised autocorrelation function $C_{ff}$ between two successive states $t-1$ and $t$ is then given by:

\begin{eqnarray}
C_{ff}(t) &\equiv& \left<f_{t-1}f_{t}\right> - \mu_{f}^{2} \\
          &   =  & \sum_{x,y}f(x)[\pi_{x}p_{xy}^{(|t|)} - \pi_{x}\pi_{y}]f(y) \nonumber \\
\end{eqnarray}

What is of most interest is the \textit{normalised autocorrelation function} $\rho_{ff}(t)$ at a given time step $t$, as normalised to the initial ($t=0$) state $\alpha$. That is to say, the normalised autocorrelation function $\rho_{ff}(t)$ provides a measure of how correlated a value $f$ (at time $t$) is to the initial distribution $\alpha$ (when $t=0$):

\begin{equation}
\rho_{ff}(t) \equiv \frac{C_{ff}(t)}{C_{ff}(0)}
\end{equation}

The normalised autocorrelation function will decay exponentially for large $t$ as the states move closer to the equilibrium distribution $\pi$. In other words, at some large $t$ (which is unknown \textit{a priori}) the state $\bm{\Theta}_t$ will be independent, or negligibly dependent on the initial state $\alpha$. Therefore it is important to know \textit{how quickly} the exponential decay of $\rho_{ff}$ is, such that I can identify where the chains have properly thermalised and settled in to $\sim{}\pi$. The \textit{exponential} autocorrelation time $\tau_{exp,f}$ provides the measure of decay and is defined as,

\begin{equation}
\tau_{exp,f} = \lim_{t\rightarrow{}\infty}{\sup}\frac{t}{-\ln{|\rho_{ff}(t)|}}
\end{equation}

\noindent{}where $\tau_{exp}$ provides the relaxation time of the system, as given by which parameter $f$ is moving slowest from $\alpha$ to $\pi$:

\begin{equation}
\tau_{exp} = \sup_{f}\tau_{exp,f}
\end{equation}

Therefore, if $t$ is large enough, the burn-in period can be estimated as a $\sim{}$few multiples of the relaxation time of the system $\tau_{exp}$. However there is some ambiguity about when $\tau_{exp}$ can be calculated, because for short chains $\tau_{exp}$ is forced to be shorter than the chain length by construction. For this reason we must sample a `sufficient' number of times before estimating the exponential autocorrelation time. Adopting conservative behaviour by default, \sick{} will initialize 200 walkers and run until $\tau_{exp} < 1/64$ the chain length. In practice this may be overkill, but this behaviour can be changed by the user.

\begin{figure*}[t!]
\includegraphics[width=\textwidth]{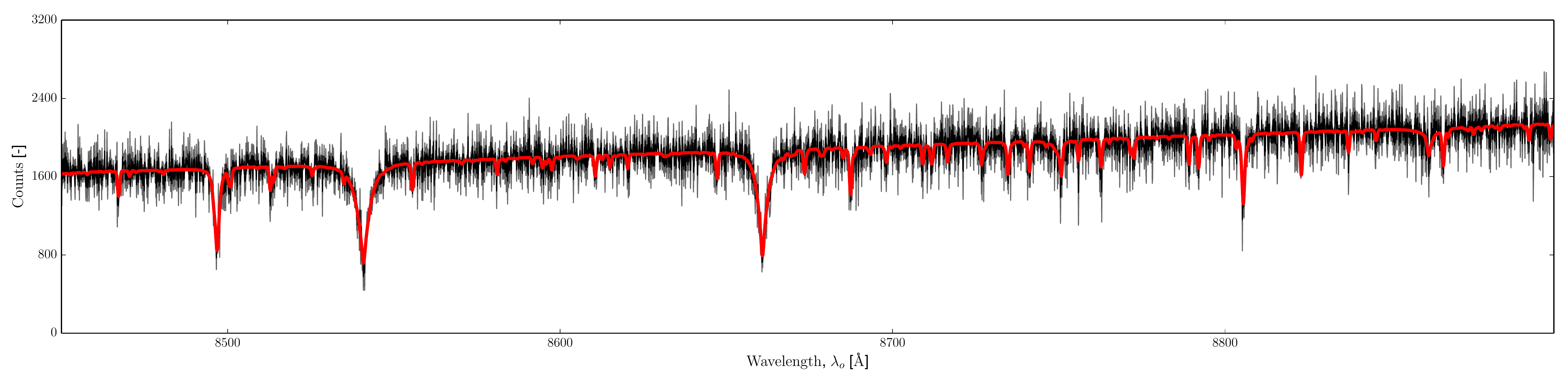}
\caption{A faux observed spectrum (black) of $\mathcal{R} \sim 20000$ and S/N 
ratio of $\sim20$ per pixel, which was used in the toy model test. The recovered maximum a posteriori model spectrum is shown in red.}

\label{fig:toy-model-projection}
\end{figure*}

I estimate $\tau_{exp}$ by first calculating the normalised autocorrelation function $\rho_{ff}$ for each parameter, using the mean position of all walkers at any time $t$. For any time $t$ I take the maximum absolute $\rho_{ff}$ for any parameter $f$: $\rho_{max} = \sup_{f}|\rho_{ff}|$. Thus, $\rho_{max}$ gives the upper limit of autocorrelation in any parameter $f$ at a given time $t$. Finally, I fit an exponential function of the form $\exp{\left(-\frac{t}{\tau_{exp}}\right)}$ to the profile $(t,\rho_{max})$ by least-squares minimisation in order to estimate $\tau_{exp}$. 

Given the estimate of $\tau_{exp}$ I discard the first 3$\times\tau_{exp}$ MCMC steps as the thermalisation phase. Using the remaining samples, I next calculate the \textit{integrated} autocorrelation time $\tau_{int}$ (using the \texttt{emcee.autocorr.integrated\_time} function), which is distinct from the \textit{exponential} autocorrelation time discussed above. The integrated autocorrelation time is defined as:

\begin{equation}
    \tau_{int,f} = \frac{1}{2} + \sum_{t=1}^{\infty}\rho_{ff}(t)
\end{equation}

Because the integrated autocorrelation time $\tau_{int,f}$ is calculated only on samples after $3\times\tau_{exp}$ (i.e., after thermalisation has occurred), it provides a measure of the statistical error in the Monte Carlo samples of parameter $\mean{f}$, and a means of determining the number of effective independent samples of $f$,

\begin{equation}
N_{eff,f} \approx \frac{N_{steps}}{2\tau_{int,f}}
\end{equation}

\noindent{}where $N_{steps}$ is the number of production (post-thermalisation) MCMC steps. The reader is referred to the excellent notes by A.~D.~Sokal\footnote{http://www.stat.unc.edu/faculty/cji/Sokal.pdf} for more details (and clear derivations) of sample estimators and autocorrelation times.

After the thermalisation regime has been identified, the default constraint for convergence in \sick{} is to have more than 100 effective independent samples in every parameter $f$. If this heuristic is not met after the first 2,000 steps, \sick{} will calculate $\tau_{exp}$, $\tau_{int}$ and $N_{eff}$ every 1,000 MCMC steps thereafter. Expert users can modify this behaviour by disabling convergence checking completely (with the \texttt{auto\_convergence} setting) in lieu of specifying the number of iterations to \texttt{burn} and \texttt{sample}, or by altering the following convergence criteria settings: \texttt{n\_tau\_exp\_as\_burn\_in}, \texttt{minimum\_effective\_independent\_samples}, \texttt{check\_convergence\_frequency}, and \texttt{minimum\_samples}, which will specify the minimum number of MCMC steps before evaluating convergence.

Once sampling is complete, \sick{} generates figures showing the normalised auto-correlation $\rho_{ff}$ in each parameter, the mean acceptance fraction at each step, all of the sampled $\bm{\Theta}$ parameters in each chain, corner plots showing marginalised posteriors for $\bm{\Theta}$ and $\bm{\theta_*}$, as well as projection (spectrum) plots that illustrate the quality of fit to the data. Furthermore, the chains and final state of every MCMC is saved by \sick{}, allowing users to resume their analysis from the most recent state, calculate additional sample estimators, or to produce supplementary post-processing figures.

\section{Toy Model}
\label{sec:toy-model}

A straightforward test of the probabilistic framework described above is to produce a noisy faux observation, and infer the model parameters $\bm{\Theta}$, given the faux data. The AMBRE public spectral library \citep{de_laverny_2012} has been employed as the grid of model intensities. I used $\sim$4000 points $\{\bm{\theta_*}\}^{\rm grid}$ in the range $4000 < T_{\rm eff} < 8000$, $0 < \log{g} < 5$, $-3 < {\rm [Fe/H]} < 0.5$ and $-0.2 < [\alpha/{\rm Fe}] < 0.8$ to train a \textit{Cannon} quadratic-in-labels  model with linear cross-terms. Stellar parameters $\{T_{\rm eff}, \log{g},{\rm [Fe/H]}, [\alpha/{\rm Fe}]\}$ of a solar-like star $\bm{\theta_*} = [5841, 4.41, -0.03, +0.01]$ were chosen, and model intensities $\bm{I_{\lambda,m}}$ were generated between $\lambda_m = [8450, 8900]$\,\AA{}. The wavelength range is quite common to many surveys and instruments: RAVE, FLAMES/GIRAFFE (HR21), Gaia RVS, AAOmega (1700I/D). For this test, the spectral resolution is most comparable to FLAMES/GIRAFFE. For the purposes of producing a faux observation, I disregard the uncertainties in producing the model intensities $\bm{\sigma_{\lambda,m}}$. In other words, $\bm{\sigma_{\lambda,m}}$ does not contribute to the \textit{quoted} noise in $\bm{\sigma_{\lambda,o}}$. However when generating model intensities for probability evaluations at run-time, $\bm{\sigma_{\lambda,m}}$ enters into the likelihood function as per Equation \ref{eq:total-variance}.

\begin{figure}[h!]
\includegraphics[width=\columnwidth]{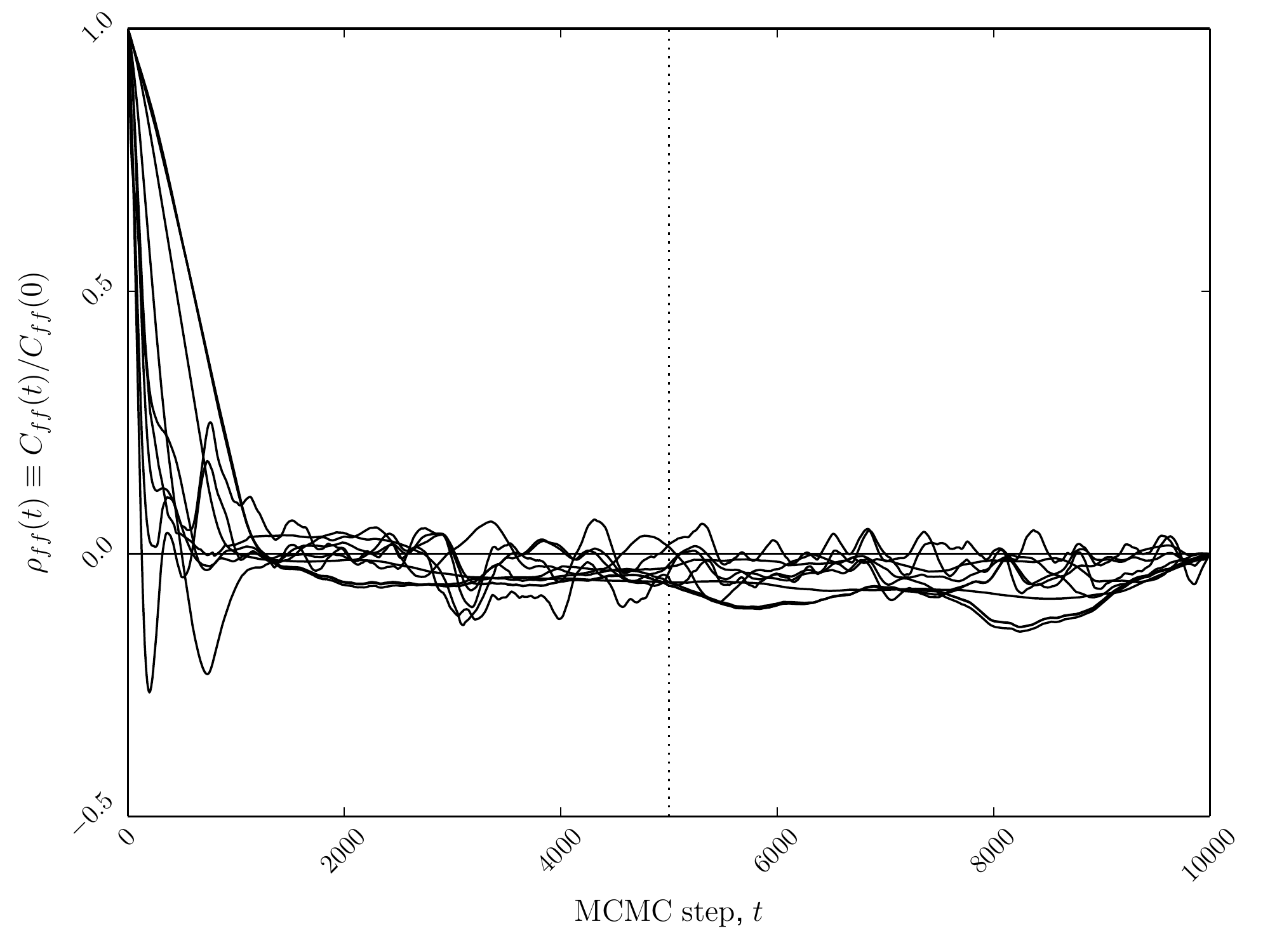}
\caption{The normalised autocorrelation function $\rho_{ff}$ for the toy model. Auto-correlations from all model parameters are shown, as calculated by the mean position of the walkers at each time step $t$. The exponential autocorrelation time is $\tau_{\exp} \sim{} 500$, demonstrating that the Markov Chains are in equilibrium (converged) well before the burn-in point, $t = 5000$ (marked).}
\label{fig:toy-model-norm-acor}
\end{figure}

A number of transformations were then applied to the intensities. The spectra were convolved to a resolving power of $\mathcal{R} \sim 20,000$ and redshifted by a random velocity drawn from $\mathcal{N}(0, 300)$\,km s$^{-1}$ before the data were binned onto a uniform spacing of $0.1$\,\AA{}. A second-order polynomial was used to represent the continuum, and noise was added to replicate a S/N ratio of $\sim{}$20 per pixel.

\begin{figure*}
\includegraphics[width=\textwidth,height=\textwidth]{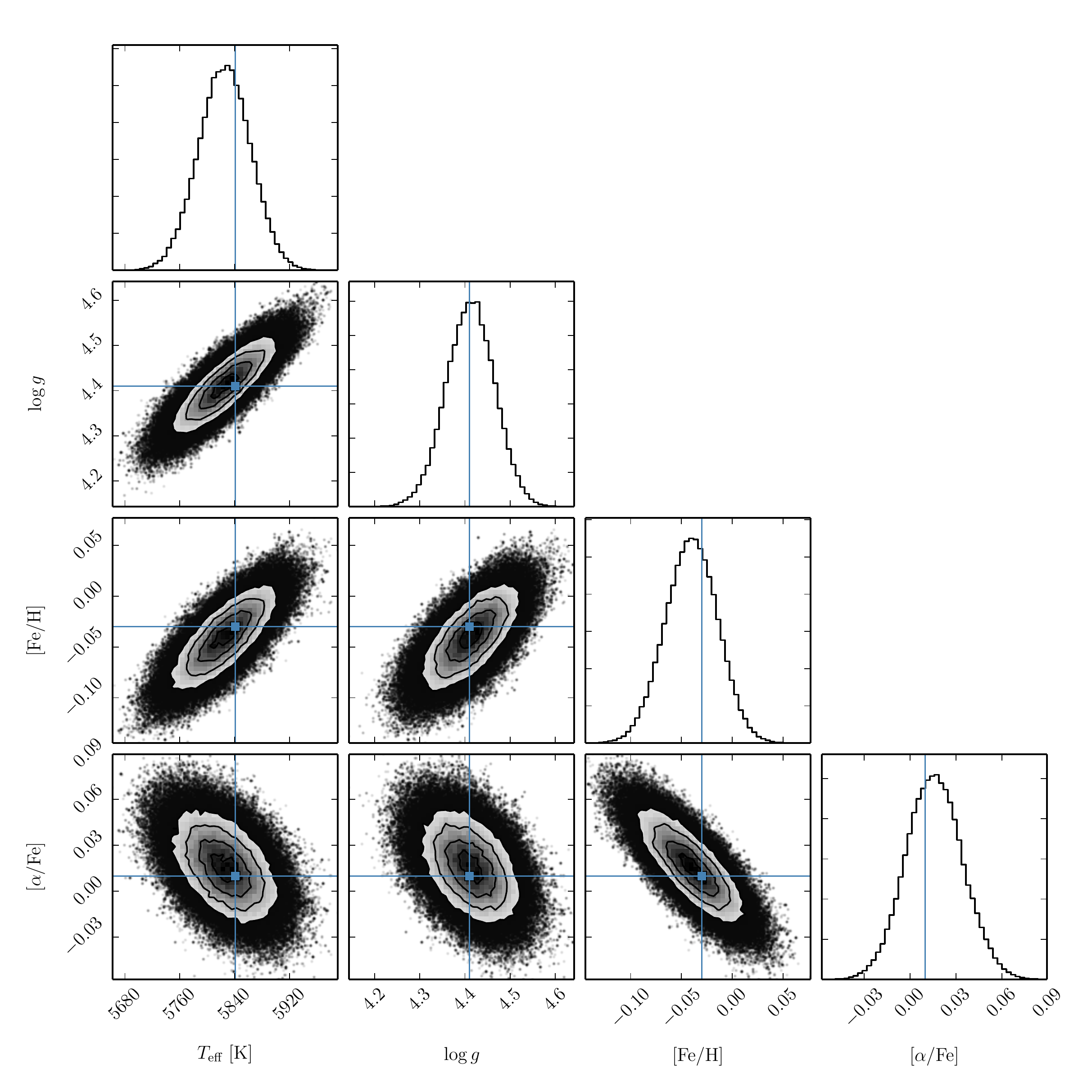}
\caption{Posterior probability distributions for all astrophysical parameters $\bm{\theta_*}$ for a faux observation with spectral resolution $\mathcal{R} \sim 20,000$ and S/N ratio $\sim{}20$\, per pixel. The posterior probability distributions of $\bm{\theta_*}$ are marginalised over $z$, $\mathcal{R}$, $\bm{c}$ and $\ln{f}$. The parameter values used to generate the data are marked in blue. This figure demonstrates the forward model described, and highlights how precise inferences of stellar parameters can be made with high-resolution spectra, even in the presence of substantial noise.}
\label{fig:toy-model-corner}
\end{figure*}

The model included the parameters $T_{\rm eff}$, $\log{g}$, [Fe/H], [$\alpha$/Fe], $z$, $\mathcal{R}$, $c_0$, $c_1$, $c_2$, and $\ln{f}$. 200 walkers sampled for 5000 steps to thermalise the sampler. In practice, this is more samples than what was necessary for this test, as evidenced by the normalised autocorrelation function $\rho_{ff}(t)$ (Figure \ref{fig:toy-model-norm-acor}). The chains were reset and another 5000 MCMC steps were performed to sample the posterior distribution. The posterior probability distributions of $\bm{\theta_*}$ are shown in Figure \ref{fig:toy-model-corner}, where blue marks the parameters used to generate the data. Table \ref{tab:toy-model} lists the maximum a posteroiri values of $\bm{\Theta}$ (which is projected to the data in Figure \ref{fig:toy-model-projection}) and the 16th and 84th percentiles of the $\bm{\Theta}$ distributions. It is clear from Figure \ref{fig:toy-model-corner} and Table \ref{tab:toy-model} that the model recovers the data-generating values $\bm{\Theta}$ very well.

\begin{table*}
\center
\caption{Model parameters and values employed for, and inferred from, the toy model.}
\label{tab:toy-model}
\begin{tabular}{llcr}
\hline
\hline
\bf{Parameter} & \bf{Description} & \bf{Data-Generating Value} & \bf{MAP Value} \\
\hline
$T_{\rm eff}$       & Effective photospheric temperature [K]                &    5841 &      $5825^{+38}_{-38}$ \\ 
$\log{}g$           & Surface gravity                                       &    4.41 &  $4.41^{+0.05}_{-0.05}$ \\ 
${\rm [Fe/H]}$      & Metallicity                                           & $-0.03$ & $-0.04^{+0.03}_{-0.03}$ \\ 
$[\alpha/{\rm Fe}]$ & $\alpha$-element enhancement                          & $+0.01$ & $+0.02^{+0.02}_{-0.02}$ \\ 
$z\cdot{}c$         & Redshift/Doppler shift [km s$^{-1}$]                  &    43.1 &    $43.2^{+0.1}_{-0.4}$ \\
$\mathcal{R}$       & Spectral resolution                                   &   20000 &  $21648^{+232}_{-1446}$ \\
$c_{0}$             & Continuum polynomial coefficient                      & $-8000$ & $-9050^{+2335}_{-2358}$ \\ 
$c_{1}$             & Continuum polynomial coefficient                      &    1.14 &  $1.39^{+0.54}_{-0.54}$ \\ 
$c_{2}$             & Continuum polynomial coefficient $(\times10^{-5})$    &    $-5$ & $-1.42^{+3.11}_{-3.14}$ \\ 
$\ln{f}$            & Logarithm of fractionally underestimated variance     &\nodata  & $-8.93^{+0.85}_{-0.73}$ \\ 
\hline
\end{tabular}
\end{table*}

\section{Utility: Atomic Diffusion in M67}
\label{sec:utility}

Although \sick{} can be used to estimate a (nearest-neighbour or numerically optimised) point estimate, I have spent considerable effort describing the dominant conceivable phenomena that may affect the data, and outlined how to incorporate those nuisance effects into a scalar-justified model. Given the additional (often considerable) computational cost implied by MCMC to marginalise over these nuisance parameters, it is reasonable to ask `\textit{Why bother?}'. Does the introduction and marginalisation of nuisance parameters actually improve our inferences on astrophysical parameters $\bm{\theta_*}$? In other words, if a full sampling of the posterior and marginalisation of nuisance parameters does not provide additional \textit{scientific} information, is it pragmatic to perform MCMC? The computational cost may not be warranted.

Here I present a suitable application that demonstrates that there \textit{is} additional information in existing public spectra which has not been fully exploited. M67 is a nearby \citep[$\sim{}$800-900\,parsec;][]{sandquist_2004, majaess, sarajedini, yakut} open cluster with a near-solar metallicity: ${{\rm [Fe/H]}= -0.04}$ to $+0.03$ \citep{hobbs_thorburn,tautvaisiene_2000,yong_2005,randich_2007,pace_2008,pasquini_2008,onehag_2011}. The age of the cluster is comparable to the Sun ($3.5-4.8$\,Gyr), and represents an excellent test-bed for stellar evolution and diffusive convection at solar metallicity \citep[e.g.,][]{vandenberg_2008}.

\begin{figure}
\includegraphics[width=\columnwidth]{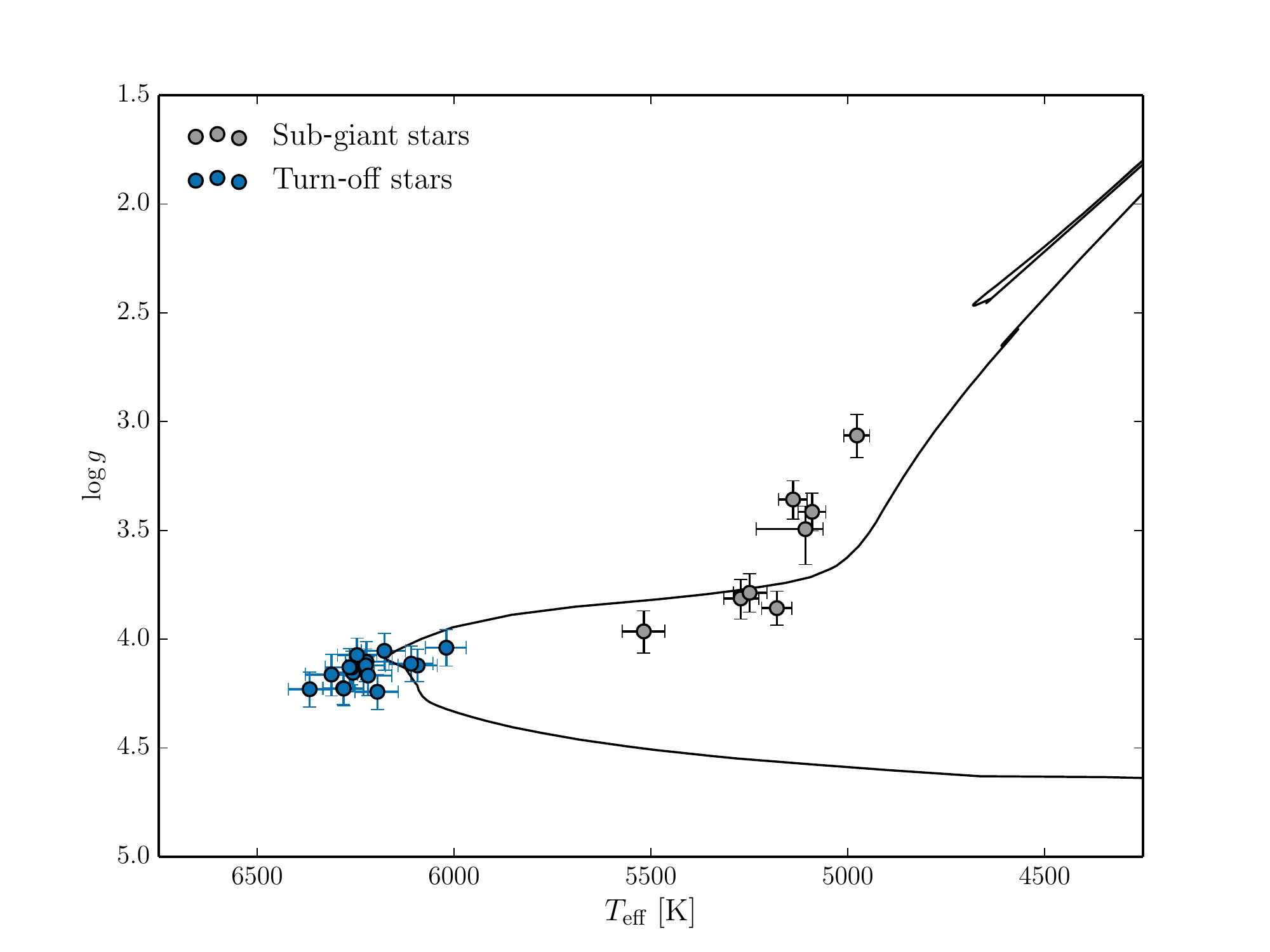}
\caption{Inferred effective temperatures and surface gravities of confirmed cluster members of M67. A single-mass 4.5 Gyr PARSEC \citep{bressan} isochrone of solar metallicity is shown to guide the eye. The discrepancy in the position of the red giant branch is a known astrophysical phenomena (see text). When accounted for, there is good overall agreement with the sequence. Samples are labelled as turn-off or sub-giant stars.}
\label{fig:m67-isochrone}
\end{figure}

Candidate stars in M67 were observed on the Australian Astronomical Telescope with the AAOmega instrument in May 2011. The 1700D grating was used in the red arm, which gives comparable wavelength coverage to the toy model in Section \ref{sec:toy-model}, but at a lower resolution of $\mathcal{R} \sim{} 10000$. I convolved the AMBRE spectral library \citep{de_laverny_2012} to this spectral resolution while keeping the high-resolution sampling. A \textit{Cannon} model with label vector [1, $T_{\rm eff}^3$, $T_{\rm eff}^2$, $\log{g}^2$, ${\rm [Fe/H]}^2$, $[\alpha/{\rm Fe}]^2$, $T_{\rm eff}\cdot{}\log{g}$, $T_{\rm eff}\cdot{}[Fe/H]$, $T_{\rm eff}\cdot{}[\alpha/{\rm Fe}]$, $\log{g}\cdot{}[\alpha/{\rm Fe}]$, ${\rm [Fe/H]}\cdot{}[\alpha/{\rm Fe}]$, $T_{\rm eff}$, $\log{g}$, ${\rm [Fe/H]}$, $[\alpha/{\rm Fe}]$] was trained across the region $4000 <= T_{\rm eff} <= 7000$, $1.0 <= \log{g} <= 5.0$, $-2.5 <= {\rm [Fe/H]} <= 0.5$, and $-0.4 <= [\alpha/{\rm Fe}] <= 0.4$ (4202 points). Once the model was trained, I used the `\texttt{sick infer}' command line (with prescribed `burn' and `sample' values, see below) to infer astrophysical parameters given the 1700D spectra. As described in Section \ref{sec:method}, the $-\ln{\mathcal{P}}$ was numerically optimised from a nearest-neighbour point estimate of the parameters $\bm{\Theta}$, before performing MCMC sampling with 200 walkers for 2000 steps in thermalisation and production. The model parameters were $\bm{\Theta} = [T_{\rm eff},\log{g},{\rm [Fe/H]},[\alpha/{\rm Fe}],z,\mathcal{R},\ln{f},c_0,c_1,c_2,p_b,v_b]$.

Cluster members were unambiguously identified from their inferred redshifts $z$. Suspected spectroscopic binaries (due to significant line broadening or resolved double-peaks in their spectrum) were discarded. The distilled sample includes 24 members, which are shown in Figure \ref{fig:m67-isochrone}, with a 4.5 Gyr solar-metallicity PARSEC \citep{bressan} isochrone to guide the eye. The sample consists of predominantly turn-off and sub-giant stars. The discrepancy with the isochrone at the giant branch is a noticeable, well-studied effect \citep[e.g.,][]{vandenberg_1983}. Indeed, the position of the red giant branch is a sensitive function of the (invoked) $\alpha$ parameter in mixing length theory. For this reason M67 has been a useful boundary condition for testing blanketed model atmospheres \citep[e.g.,][]{vandenberg_2008}. Without this condition, the position of the red giant branch in 4.5 Gyr solar isochrones tends redwards (e.g., cooler temperatures), to the same degree that I find in Figure \ref{fig:m67-isochrone}. The uncertainties are sufficiently large that I cannot precisely distinguish between stars that have passed the turn-off point, but the overall agreement with the sequence is satisfactory.

\begin{figure}
\includegraphics[width=\columnwidth]{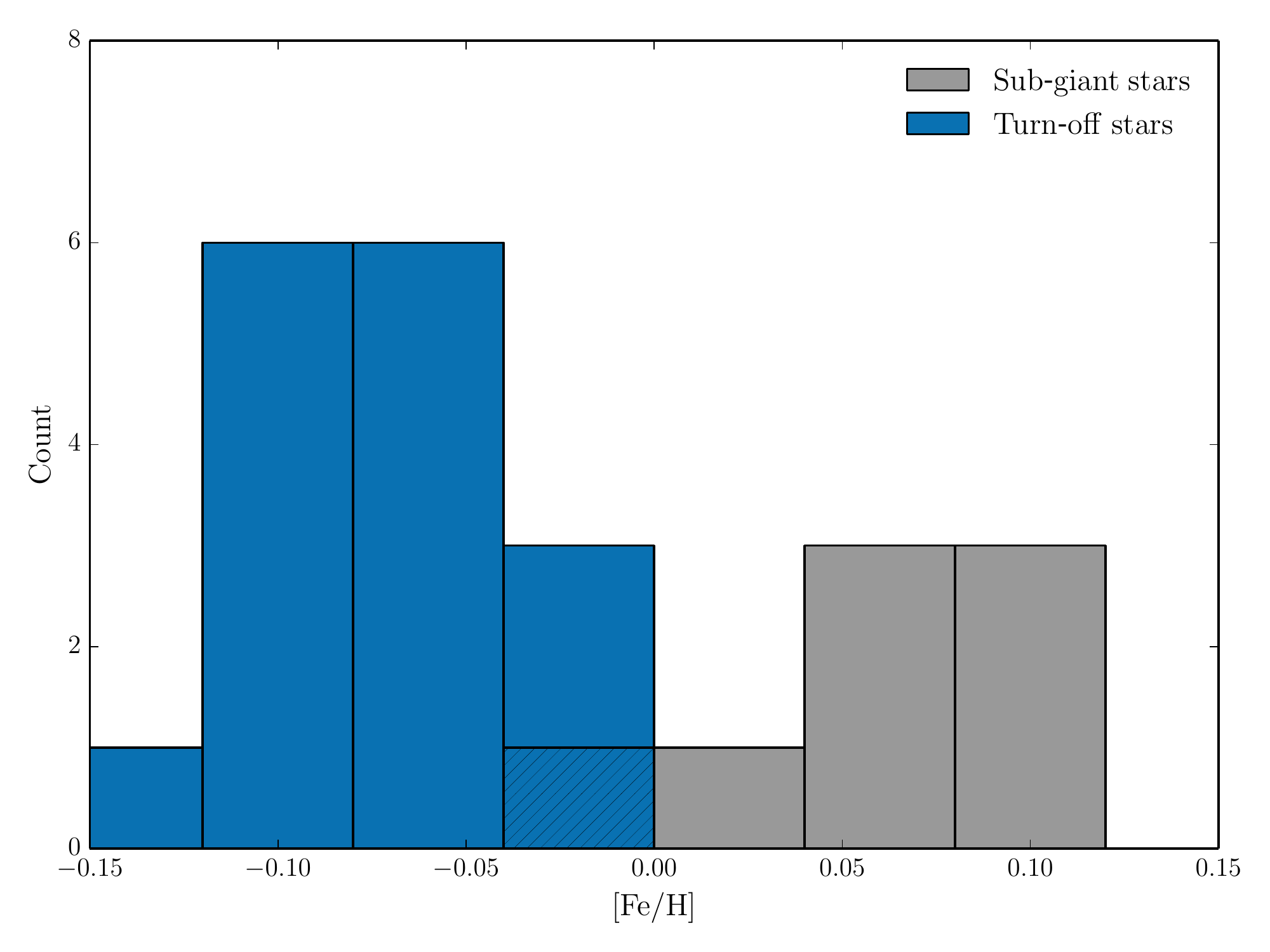}
\caption{A histogram showing the maximum a posteriori metallicity for confirmed M67 stars,  colored by their evolutionary stage as per Figure \ref{fig:m67-isochrone}. As a whole, the sub-giant stars exhibit slightly ($+0.05$\,dex, when individual uncertainties in [Fe/H] are considered) higher metallicities than the turn-off stars, a result previously identified from high-resolution spectra by \citep{onehag_2014} and attributed to atomic diffusion.}
\label{fig:m67-metallicities}
\end{figure}

If I assume that the underlying metallicity distribution of M67 is a Gaussian, and account for the uncertainties in [Fe/H] for each star, from all cluster members I find the metallicity of M67 to be [Fe/H] $= -0.06$. This measurement is in \textit{reasonable} agreement with existing studies that place the metallicity of M67 between $-0.04$ and $+0.03$. However, Figure \ref{fig:m67-isochrone} shows the sample is two-thirds dominated by turn-off stars. In Figure \ref{fig:m67-metallicities} I show the inferred maximum a posteriori metallicity for all M67 stars, binned to 0.05\,dex increments (i.e., roughly equivalent to the uncertainties in individual measurements). The turn-off and sub-giant stars are colored in the same way as Figure \ref{fig:m67-isochrone}, and the hatching represents an overlap between the two distributions. As a whole, the turn-off stars are more metal-poor than the sub-giant stars. If the uncertainties in [Fe/H] for each star are considered, and assume the underlying metallicity distribution in the sub-giant and turn-off sample are normally distributed, I find the metallicity of turn-off stars to be [Fe/H]$ = -0.07$, and [Fe/H]$ = -0.02$ for the sub-giant stars.

The difference in metallicities between sub-giant and turn-off stars in M67 is not a new result. \citet{onehag_2014} performed a differential study of sub-giant and turn-off stars in M67 with respect to the solar twin M67-1194. Their data were obtained from the UVES instrument on the VLT, providing $\mathcal{R} \sim 47,000$ and S/N $\sim{}$ 150 per (binned) pixel. This data quality permitted the determination of individual chemical abundances in a strictly differential sense.
It is important to remember that the metallicity difference I find is probably a convolution of multiple elements within this wavelength range (Fe, Ti, and Ni). With this caveat in mind -- and although \citet{onehag_2014} cover a slightly different temperature range -- the effect I find is the same: heavy-element abundances (including Fe) in M67 are found to be reduced in the hotter stars and dwarfs by typically $\leq0.05$\,dex, as compared to the abundances of the sub-giants. Thus, with an objective characterisation of (most) dominant phenomena that will affect the observations, subtle astrophysical phenomena can be inferred from lower-resolution spectra than what is typically considered. This example application shows, to a large extent, that existing stellar spectroscopic data are sufficiently high-quality that the standard of our results will be dominated by our analysis methods. For this reason, a move towards generative models in stellar spectroscopy is essential.

\section{Discussion}
\label{sec:discussion}

Here I discuss some future applications that for \sick{}, an important potential caveat, and outline planned near-term improvements for the code.

\subsection{Further Applications}

As previously discussed, \sick{} is agnostic about wavelength coverage, resolving power, or binning of the observed data. This generality allows for an extremely high-resolution grid to be used for many applications of lower resolutions, as long as the wavelengths are covered. For example, the same high-resolution library (observed or synthetic) can be used for surveys of high-resolution (e.g., APOGEE), low-resolution (e.g., SEGUE, Gaia RVS or Gaia BP/RP\footnote{Indeed, \sick{} performed excellently in a blind test during the 3rd \textit{Gaia Challenge Workshop}, slightly outperforming the current BP/RP analysis method for metal-poor stars: \href{http://astrowiki.ph.surrey.ac.uk/dokuwiki/doku.php?id=tests:astropars:challenge3}{\texttt{http://astrowiki.ph.surrey.ac.uk/dokuwiki/doku.php?id=tes\\
ts:astropars:challenge3}}} -- where $\mathcal{R} \sim{} 120$), placing stars from all surveys on the same, self-consistent scale.

The flexibility in methodology (e.g., \texttt{estimate}, \texttt{optimise}, \texttt{infer}) also allows for quick analyses over extremely large data sets (e.g., LAMOST) to identify superlative objects with high scientific impact (e.g., ultra metal-poor or hyper-velocity stars). Similarly, large collections of point estimates can be sufficient to identify and quantify substructure in the Milky Way halo. Alternatively, sampling posterior probability distributions for a reasonable sample of cluster stars may reveal subtle abundance variations, helping to untangle the `multiple population' scenario \citep[e.g., see][and references therein]{carretta} or understand the effects of atomic diffusion in clusters of different ages and metallicities. \\

\subsection{Caveats}
Although there are substantial scientific applications for the tool presented here, there is an obvious caveat that requires attention. The examples here have focused on producing model intensities $\bm{I_{\lambda,m}}$ from at most four dimensions. As the number of dimensions increase -- for example, to include individual chemical abundances -- the \textit{curse of dimensionality} will quickly become relevant. The computational complexity in producing $\bm{I_{\lambda,m}}$ scales quickly, such that a linear interpolation model will become absolutely unsuitable in higher dimensions. On the other hand, there are some tricks that can be introduced for a \textit{Cannon}-like approach in high-dimensionality for chemical abundances \citep[e.g., see ][]{Ness_2016,Casey_2016}.  Alternatively, compounding two suitable models may be a reasonable scenario: one for the determination of stellar parameters, and another that produces spectra of individual elemental abundances (by some means, \textit{Cannon}-like or synthesised) for some small wavelength region and stellar parameters $\bm{\theta_*}$. The abundance of an individual element can be marginalised over all possible $\bm{\theta_*}$. The implied assumption here is that the individual chemical abundances do not have a substantial effect on the overall stellar parameters. It is tempting to assert that this is an unwarranted generalisation. However this is an implied assumption used in the production of model photosphere, since the photospheres themselves are calculated with a given chemical composition of individual elements. Although there are potential ways to deal with higher dimensionality (in stellar applications), \textit{extreme} care must be made in scaling the intensity-generating methods described here.

There are a number of caveats that users must be cognisant of even for models with low-dimensionality.  These apply to the data and the model employed. For example, this framework is most suitable for one-dimensional extracted spectra; it is beyond the scope of this work to forward model two-dimensional images.  Likewise, \sick{} may be sub-optimal for echelle spectra with many orders due to the high number of $\{\bm{c}\}$ coefficients that are physically related.

Even within a single order, there are strong assumptions made about the line spread function.  It is assumed that the resolution scales (at most) linearly with wavelength, which can be shown to be knowingly incorrect for most spectrographs. The effect of this assumption will usually be small (a small second order resolution term is probably warranted), but it should be known.

In the work presented here, flux noise is assumed to be Gaussian. Similarly the noise can only assumed to be under-estimated (not over-estimated), and no framework has been presented to account for correlated noise between neighbouring pixels \citep[e.g., ][]{czekala}. It is also important to note that the treatment of outliers in this work may be overkill: rejecting highly discrepant pixels may be sufficient. In short, the user should be extremely familiar with the model description they set out for the data, and the limitations thereof. Improvements to the code that help resolve these existing caveats are welcomed in the form of pull requests through GitHub\footnote{github.com/andycasey/sick}.

\section{Conclusion}
\label{sec:conclusion}

I have presented a flexible probabilistic code to forward model spectroscopic data. The generative model approach described here has a number 
of advantages over previously published techniques. Preparatory and subjective
decisions (e.g., redshift and placement of continuum) are objectively treated 
within a scalar-justified mathematical model, allowing for a credible assessment of uncertainties in astrophysical parameters. Almost all previously published techniques 
have treated these processes separately, thereby increasing biases in their results and generally mis-characterising the uncertainties in astrophysical parameters.

The simultaneous incorporation of continuum, redshift, convolution and resampling leads to remarkable improvements in both accuracy and precision. While the examples 
presented here have focused on stellar spectra, the code is ambivalent about 
\textit{what} the astrophysical parameters describe: the framework can be easily 
used for any kind of quantifiable astrophysical process. The code is MIT-licensed 
(freely distributable, open source) and has an extensive automatic testing suite. Complete documentation is available online, 
which includes a number of additional examples and tutorials on analysing data from well-known stellar surveys (e.g., SEGUE, LAMOST, APOGEE).

Great effort has been made to ensure the code is easy to use, allowing users to 
obtain precise inferences with little effort. I strongly encourage the use of this 
software for existing and future spectral data. With the sheer volume of 
high-quality of spectra available, astronomers must begin to adopt objective, 
generative models for their data. Subtle astrophysical processes can only be 
discovered and understood with the proper characterisation of uncertainties 
afforded by generative models.

\acknowledgements
I am pleased to thank Matt Auger, Sergey Koposov, Melissa Ness, Jarryd Page, Jason Sanders, and Kevin Schlaufman. I am particularly
indebted to Brian Schmidt for introducing me to Bayesian statistics and generative models (far too late,
I might add), and to the referees for their constructive and direct criticism,
which improved the methodology of the code and overall clarity of this paper.
This research has made extensive use of NASA's Astrophysics Data System 
Bibliographic Services, the 
Coveralls continuous integration service, GitHub, and the \textsc{triangle.py} 
code \citep{triangle.py}. The author recognises support through the European 
Research Council grant 320360: The Gaia-ESO Milky Way Survey. 
The source code for this research is distributed using \textsc{git}, and is hosted online at 
GitHub. Suggestions for improvements or unexpected behaviour can be reported through GitHub issues, and code contributions are welcomed in the form of pull requests.

\bibliographystyle{apj}
\bibliography{biblio}

\end{document}